\documentclass[conference]{IEEEtran}

\usepackage[square,numbers,sort&compress]{natbib}
\usepackage{hyperref} %
\usepackage%
{hyperref} %
\hypersetup{
    colorlinks,
    linkcolor={blue!80!black},%
    citecolor={green!50!black},
    urlcolor={blue!80!black}
}

\usepackage{amsthm}
\usepackage{amsmath}
\usepackage{mathrsfs}
\usepackage{amssymb} 
\usepackage{stackrel}
\usepackage{mathtools}
\usepackage{nicefrac}

\usepackage{physics}

\usepackage{verbatim}
\usepackage{enumerate}

\usepackage[T1]{fontenc}
\usepackage{bbm} 
\usepackage{bbding}
\usepackage{keystroke}
\usepackage{pifont}
\usepackage{relsize} %

\usepackage{graphicx}
\usepackage{xcolor}
\usepackage{tikz}
\usetikzlibrary{shapes, arrows.meta, positioning}
\usepackage{pgfplots}
\pgfplotsset{compat=1.18}
\usetikzlibrary{calc}
\usepackage{tabularx}
\usetikzlibrary{arrows,shapes}

\usepackage{balance}

\renewcommand{\trace}{\mathrm{Tr}}

\newcommand{\identity}{\mathbbm{1}}

\theoremstyle{remark}	
\theoremstyle{remark}	
\theoremstyle{remark}	
\theoremstyle{remark}	
\theoremstyle{remark} 
\theoremstyle{remark} 
\theoremstyle{remark}

\title{Quantum Teleportation  Over a Noisy Relay Channel}

\author{\IEEEauthorblockN{Yigal Ilin}
\IEEEauthorblockA{\textit{Classiq Quantum Computing} \\
\tt yigal@classiq.io}
\and
\IEEEauthorblockN{Uzi Pereg}
\IEEEauthorblockA{\textit{ECE Department} \&\\  \textit{Helen Diller Quantum Center,} \\
\textit{Technion}\\
\tt uzipereg@technion.ac.il}
}

\date{\today}

\begin{document}

\maketitle

\begin{abstract}
Quantum teleportation and Bell measurements are considered in a noisy relay setting. We introduce two quantum relay-channel models with closed-form capacity formulas, motivated by the canonical quantum communication protocols of teleportation and superdense coding. In both models, the sender transmits a qubit, while the relay observes side information about the Pauli errors that occur in the channel and communicates with the receiver through an orthogonal rate-limited link. We develop a compress-forward coding scheme in which the relay compresses the Pauli-error sequence into bins and sends the bin index to the receiver. The receiver uses this partial error information to reduce the effective channel noise before quantum decoding. We show that this strategy is optimal by establishing matching converse bounds for both models. 
\end{abstract}

\maketitle
\section{Introduction}
Quantum teleportation and superdense coding are two of the most fundamental communication protocols in quantum information theory. %
Teleportation enables the transmission of unknown quantum states by combining shared entanglement with classical communication, whereas superdense coding exploits shared entanglement to increase the classical communication rate. %
These protocols can also be viewed %
as components of larger communication networks \cite{BFSDBFJ:20b}.

Relay channels provide a basic information-theoretic model for cooperative communication.  The fundamental building block of communication via a relay is illustrated %
below:
\begin{figure}[hb]
    \centering
    
\begin{tikzpicture}[scale=0.5,
    node distance=1.5cm and 1.8cm,
    >=latex,
    sum/.style={draw, circle, inner sep=0pt, minimum size=2mm},
    box/.style={draw, align=center}
]

\node [sum] at (3,3) (v1) {};
\node [sum] at (9,3) (v2) {};
\node [sum] at (6,5) (v3) {};    

    \begin{scope}[line width=1pt]

             \draw[->,thin] (v1) node[left,xshift=-0.25cm,yshift=0.025cm]{ {\small sender}} -- (v2) node[right,xshift=0.25cm,yshift=0.025cm]{ {\small destination}};
             \draw[->,thin] (v3)  -- (v2) node[right,xshift=0.25cm,yshift=0.025cm]{};
		\draw[->,thin] (v1) -- (v3) node[right,xshift=0.25cm,yshift=0.075cm]{ {\small relay}};

 \end{scope}

\end{tikzpicture}

    \caption{Relay channel schematic.}
    \label{fig:Relay_schematic}
\end{figure}
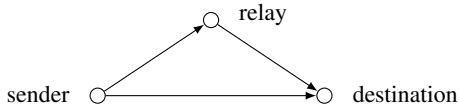

\noindent
where the relay is an intermediate  node that helps the transmission of information from the sender to the destination receiver.  
 In the classical setting, the relay channel has been studied extensively \cite{KramerGastparGupta:05p}. %
Of particular relevance here is the \emph{primitive} relay channel \cite{Kim:07c}, in which the relay assists the destination through an orthogonal rate-limited link. Despite substantial progress on achievable schemes, converses, and special classes of primitive relay channels \cite{TandomUlukus:10c,MondelliHassaniUrbanke,GohariNair:22p}, the capacity remains unknown in general. %

Quantum settings of network communication have recently received increasing attention \cite{SreekumarBertaHircheCheng:25c,salek2025three,NaturPereg:25p,BernardPadakandla:26a,ShenBloch:26a,yao2026unlocking,MaiLaBattouAmlouNunn:26a}.
Quantum relay channels provide a natural framework for studying communication in quantum networks, and play a central role in  overcoming transmission losses and enabling long-distance quantum communication
\cite{SavovWildeVu:12c,ShiShiPengGuoYiLee:12p,BochCaiDeppe:15p,pirandola2019end,Pereg:25p,IlinPereg:25c}.  %
 In a previous work, the last author \cite{Pereg:25p} introduced a fully quantum relay-channel model and established achievable strategies for the transmission of classical messages. %
 More recently, we extended this framework to relaying of entanglement %
 through a decode-and-forward coding strategy \cite{IlinPereg:25c}.

In this paper, we introduce two quantum relay-channel models with  closed-form capacity formulas. 
In both models, the sender transmits a qubit, while
the relay receives side information on the Pauli errors that occur in the channel. See Figures~\ref{Figure:Dephasing_relay}-\ref{Figure:Depolarizing_relay}. The relay then sends a message through an orthogonal link. Thereby, the receiver obtains a distorted qubit state along with a classical message from the relay.   
The structure of these models is closely related to the two canonical quantum communication protocols of teleportation and superdense coding.

\begin{figure}[tb]
    \centering
    
\begin{tikzpicture}[
    node distance=1.5cm and 1.8cm,
    >=latex,
    sum/.style={draw, circle, inner sep=0pt, minimum size=6mm},
    box/.style={draw, inner sep=3pt, align=center}
]

    \node (X) {$A$};
    \node [box, right=1cm of X] (PZ) {$\mathsf{Z}^\ell$};
    
    \node [above=1cm of PZ] (ell) {$\ell \sim 
    \text{Bern}(q_{\mathsf Z})$};
    
    \node [box, right=1cm of ell] (relay) {relay};

    \node [right=1.2cm of relay] (B0) {};
    
    \node [right=1.2cm of B0] (add3) {};
    \node [above=0.8cm of add3] (N) {};

    \node [right=4.5cm of PZ] (B) {$B$};

    \draw [->,red] (X) -- (PZ);
    \draw [->,double] (ell) -- (PZ);
    \draw [->,double] (ell) -- (relay);

    \draw [->,double] (relay) -- node[midway,above] {$R_0$} (B);

    \draw [->,blue] (PZ) -- (B);

\end{tikzpicture}

    \caption{Dephasing relay channel}
    \label{Figure:Dephasing_relay}
\end{figure}
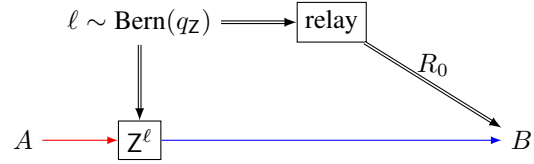
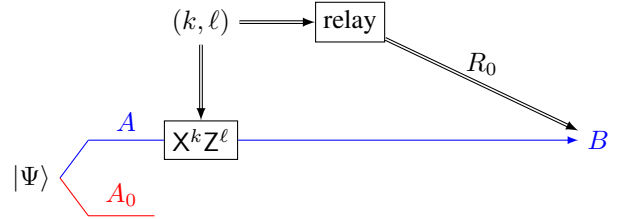
\begin{figure}[t]
    \centering
    
\begin{tikzpicture}[
    node distance=1.5cm and 1.8cm,
    >=latex,
    sum/.style={draw, circle, inner sep=0pt, minimum size=6mm},
    box/.style={draw, inner sep=3pt, align=center}
]

    \node (X) {};
    \node [below=1cm of X] (A_0) {};
    \node [below=0.5cm of X] (Psi) {};
    \node [box, right=1cm of X] (PZ) {$\mathsf{X}^k \mathsf{Z}^\ell$};

    \node [above=1cm of PZ] (ell) {$(k,\ell)$};
    
    \node [box, right=1cm of ell] (relay) {relay};

    \node [right=1.2cm of relay] (B0) {};
    
    \node [right=1.2cm of B0] (add3) {};
    \node [above=0.8cm of add3] (N) {};

    \node [right=4.5cm of PZ,blue] (B) {$B$};

    \draw [blue] (X) -- node[midway,above]{$A$} (PZ);
    \draw [blue] ($(X)+(0.125,0)$) --  ($(X)+(-0.25,-0.5)$) node[left,black]{$\ket\Psi$}; 
    \draw [red]   ($(X)+(-0.25,-0.5)$)-- ($(X)+(0.125,-1)$)-- node[midway,above]{$A_0$} ($(X)+(1,-1)$) ;
    \draw [->,double] (ell) -- (PZ);
    \draw [->,double] (ell) -- (relay);

    \draw [->,double] (relay) -- node[midway,above] {$R_0$} (B);

    \draw [->,blue] (PZ) -- (B);

\end{tikzpicture}

    \caption{Depolarizing relay channel}
    \label{Figure:Depolarizing_relay}
\end{figure}

In the first model, the destination receives a phase-flipped quantum state while the relay observes the corresponding error flag, as shown in Figure~\ref{Figure:Dephasing_relay}. This channel naturally models quantum teleportation with imperfect correction communication, where the relay must convey the missing Pauli correction through a rate-limited communication link. We derive a closed-form expression for the quantum capacity of this relay channel.

In the second model, bit- and phase-flip errors transform the transmitted Bell state, while the relay observes side information about the Pauli error, as shown in Figure~\ref{Figure:Depolarizing_relay}. The destination performs Bell-state discrimination using rate-limited relay information, giving a noisy relay-channel interpretation of superdense coding. We derive the entanglement-assisted capacity in closed form.

The optimal strategy is a quantum analogue of compress-forward: the relay compresses the error sequence into bins, enabling the destination to partially remove the channel noise before quantum decoding.
 Unlike the decode-forward coding schemes in \cite{SavovWildeVu:12c,ShiShiPengGuoYiLee:12p,BochCaiDeppe:15p,pirandola2019end,Pereg:25p,IlinPereg:25c}, the relay does not decode information at all.
We derive matching converse bounds for both models, thereby establishing the optimality of the proposed compress-forward strategy and obtaining the corresponding capacities in closed form. The resulting capacities as functions of the noise parameters are depicted in
Figures~\ref{Figure:Dephasing_Capacity}-\ref{Figure:Depolarizing_Capacity}. 

The remainder of the paper is organized as follows. Section~\ref{Section:Results} presents the channel models and the corresponding capacity results. Section~\ref{Section:Achievability} develops the compress-forward coding schemes and establishes achievability. Section~\ref{Section:Converse} provides the converse proof. Section~\ref{Section:Summary} concludes with a summary and discussion.

\section{Main Results}
\label{Section:Results}
 We use standard qunatum information notation~\cite{Wilde:17b}.
A quantum state is described by a density operator $\rho$ 
 on a finite-dimensional %
 Hilbert space $\mathcal{H}$. %
The quantum entropy is defined  %
by 
$%
H(\rho) \equiv -\trace[ \rho\log(\rho) ]
$.
Then, the quantum mutual information, conditional entropy, and
coherent information are given by
$%
I(A;B)_\rho=H(\rho_A)+H(\rho_B)-H(\rho_{AB}) %
$, 
$H(A|B)_{\rho}=H(\rho_{AB})-H(\rho_B)$, and
$%
I(A\rangle B)_\rho=-H(A|B)_\rho %
$, respectively. %
The Pauli operators are denoted by $\mathsf{X}$, $\mathsf{Y}$, and $\mathsf{Z}$, in sans-serif font.

\subsection{Teleportation Over a Relay Channel}
\label{Subsection:Teleportation_Channel}
In the \emph{noiseless} teleportation protocol, Alice holds a qubit in a quantum state $\sigma_A$ that she would like to send to Bob. 
Furthermore, Alice and Bob are provided with an entangled pair $\ket{\Phi}_{A_1 B}$.
As Alice performs a Bell measurement, she obtains an outcome
$(k,\ell)\in \{0,1\}^2$, and Bob's state becomes
$\rho_B^{(k,\ell)}\equiv\mathsf{X}^k \mathsf{Z}^\ell \sigma_A \mathsf{Z}^\ell \mathsf{X}^k$.
Alice sends the classical outcome to Bob, and then he recovers the state $\sigma_A$ by applying  bit- and phase-flip corrections.

Now, suppose that Alice sends the phase information $\ell$ through a relay, which in turn, sends the information to Bob through an orthogonal noisy classical link of capacity $R_0$ (or, equivalently,  a noiseless rate-limited link), where $R_0\in [0,2]$. That is, while the sender transmits $n$ qubits, the relay can only transmit $nR_0$ classical bits to the receiver. 
The channel is illustrated in Figure~\ref{Figure:Dephasing_relay}, where classical links are indicated as double lines. 
We show that this restriction reduces the quantum capacity to
\begin{align}
    C_{\text{Q}}=\min\left\{1\,,\; 1-h_2(q_\mathsf{Z})+R_0\right\}
\end{align}
for $\ell\sim\text{Bernoulli}(q_{\mathsf{Z}})$, where
$q_{\mathsf{Z}}\in \left( 0,\frac{1}{2} \right)$
and
$h_2(q)=-(1-q)\log(1-q)-q\log q$ is the binary entropy function.
See Figure~\ref{Figure:Dephasing_Capacity}.
If the link from the relay to the destination receiver can accommodate a sufficiently high rate, $\mathsf{R}_0\geq h_2(q_\mathsf{Z})$, then 
$C_{\text{Q}}=1$ (qubit per channel use).

We provide the achievability and converse proofs in Subsections~\ref{Subsection:Dephasing_Achievability} and 
\ref{Subsection:Dephasing_Converse}, respectively. 

\begin{figure}
    \centering

  \begin{tikzpicture}
    \begin{axis}[
        width=5.75cm,       %
        height=4.5cm,      %
      xlabel={$q_{\mathsf Z}$},
      ylabel={$C_{\text{Q}}$},
      ylabel style={rotate=-90}, %
      xmin=0, xmax=0.5,
      ymin=0, ymax=1.1,
      xtick distance=0.1,
      ytick distance=0.25,
      grid=both,
      grid style={dashed, gray!30},
      axis lines=left,
      declare function={
        R0 = 0.4; %
        h2(\x) = (\x <= 0 ? 0 : -\x*(ln(\x)/ln(2)) - (1-\x)*(ln(1-\x)/ln(2)));
        f(\x) = min(1, 1 - h2(\x) + R0);
      }
    ]
      \addplot[
        domain=0.00001:0.5,
        samples=300,
        thick,
        blue
      ] {f(x)};
    \end{axis}
  \end{tikzpicture}

    \caption{Quantum capacity for the dephasing relay channel as a function of the dephasing probability, when the relay is limited to $R_0=0.4$ (bits per channel use). }
    \label{Figure:Dephasing_Capacity}
\end{figure}
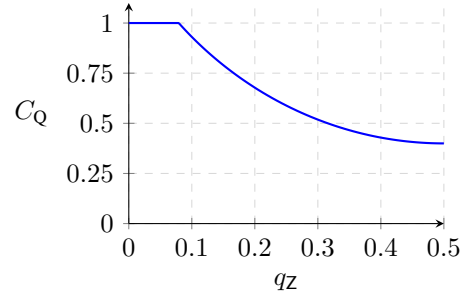
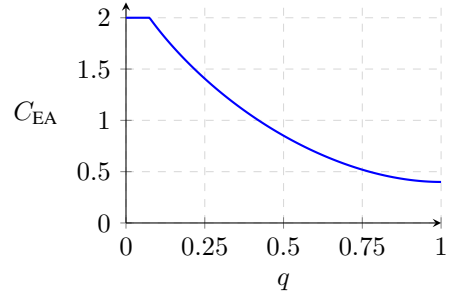
\begin{figure}
    \centering

  \begin{tikzpicture}
    \begin{axis}[
        width=5.75cm,       %
        height=4.5cm,      %
      xlabel={$q$},
      ylabel={$C_{\text{EA}}$},
      ylabel style={rotate=-90}, %
      xmin=0, xmax=1,
      ymin=0, ymax=2.15,
      xtick distance=0.25,
      ytick distance=0.5,
      grid=both,
      grid style={dashed, gray!30},
      axis lines=left,
      declare function={
        R0 = 0.4; %
        h4q(\x) = (\x <= 0 ? 0 : -(1-0.75*\x)*(ln(1-0.75*\x)/ln(2)) - 0.75*\x*(ln(\x/4)/ln(2)));
        f(\x) = min(2, 2 - h4q(\x) + R0);
      }
    ]
      \addplot[
        domain=0.00001:1,
        samples=300,
        thick,
        blue
      ] {f(x)};
    \end{axis}
  \end{tikzpicture}

    \caption{Entanglement-assisted classical capacity for the depolarizing relay channel as a function of the probability of depolarization, when the relay is limited to $R_0=0.4$ (bits per channel use). }
    \label{Figure:Depolarizing_Capacity}
\end{figure}

\subsection{Noisy Bell Measurement}
\label{Subsection:Bell_Relay}
We now move to our second relay model. We first provide the motivation. 
Consider an entanglement-based protocol that results in a  state $\ket{\Psi}_{A_0 A}$. See diagram in  Figure~\ref{Figure:Depolarizing_relay}.
Suppose that Bob would like to perform a measurement in the Bell basis, 
$\left\{\ket{\Phi^{(i,j)}}\equiv (\mathsf{X}^i \mathsf{Z}^j\otimes \identity)\ket{\Phi}\right\}_{i,j\in\{0,1\}}$.
This is particularly relevant to the superdense coding protocol.
We consider a noisy measurement that produces $\ket{\Phi^{(i\oplus k,j\oplus \ell)}}$ instead, while the relay provides rate-limited side information on $(k,\ell)$ to the receiver.

The classical capacity depends on the noise model. In particular, if the Pauli errors are independent, then the entanglement-assisted capacity is given by 
\begin{align}
    C_{\text{EA}}=\min\left\{2\,,\; 2-h_2(q_\mathsf{X})-h_2(q_\mathsf{Z})+\mathsf{R}_0\right\}
\end{align}
for $k\sim\text{Bernoulli}(q_{\mathsf{X}})$ and 
 $\ell\sim\text{Bernoulli}(q_{\mathsf{Z}})$.

Whereas, in the depolarizing noise model,
\begin{align*}
    &\mathcal{M}_{A\to B}(\rho)= (1-q)\rho+q\frac{\identity}{2}\\
    &=\left(1-\frac{3q}{4} \right)\rho+
    \frac{q}{4}\left( \mathsf{X}\rho\mathsf{X}+\mathsf{Z}\rho\mathsf{Z}+\mathsf{X}\mathsf{Z}\rho \mathsf{Z}\mathsf{X}\right)
\end{align*}
hence,
$(k,\ell)\sim\left(1-\frac{3q}{4}\,,\;\frac{q}{4}\,,\;\frac{q}{4}\,,\;\frac{q}{4} \right)$. 
We show that the classical entanglement-assisted capacity in this case is 
\begin{align}
    C_{\text{EA}}=\min\left\{2\,,\; 2-h_4\left(\frac{3q}{4}\right)+\mathsf{R}_0\right\}
\end{align}
where $h_d(t)=-\left(1-t\right)\log\left(1-t\right)-t\log(\frac{t}{d-1})$.  
For $q=1$, i.e., a completely-depolarizing channel, we have $C_{\text{EA}}=\mathsf{R}_0$.
Note that without the relay, the capacity is zero in this case. 

If the relay link has sufficiently high rate, then the receiver can correct the errors in full and the classical capacity becomes 
$C_{\text{EA}}=2$ (bits per qubit channel use), as in  noiseless superdense coding.

We give the achievability and converse proofs in Subsections~\ref{Subsection:EA_Achievability} and 
\ref{Subsection:EA_Converse}, respectively. 

\section{Achievability Proof}
\label{Section:Achievability}
We introduce a compress-forward coding scheme, where the relay sends a compressed representation that captures part of the error information, and forwards this representation to the receiver. 

\subsection{Sending Quantum Information}
\label{Subsection:Dephasing_Achievability}
Consider the quantum relay channel in Subsection~\ref{Subsection:Teleportation_Channel}.
First, we note that if $R_0\geq h_2(q_\mathsf{Z})$, then the result follows immediately, since the relay can send the error sequence $\ell^n$ to the receiver by lossless compression, and then the receiver can cancel out the error. 
This results in a noiseless transmission of $1$ qubit per channel use.

Henceforth, assume $R_0< h_2(q_\mathsf{Z})$.
Let $\alpha\in \left[0,\frac{1}{2}\right)$   such that 
\begin{align}
    h_2(\alpha)= h_2(q_\mathsf{Z})-R_0+\varepsilon 
    \label{Equation:h2alpha}
\end{align}
where $\varepsilon>0$ is arbitrarily small. 
Then, for sufficiently small $\varepsilon$, we have  $0\leq\alpha< q_\mathsf{Z}$.
It follows that for every 
$L\sim \text{Bernoulli}(q_{\mathsf{Z}})$ and 
\begin{subequations}
\label{Equations:T_V_L}
  \begin{align}
T\sim \text{Bernoulli}(\alpha) \,,
\end{align}
there exists   
$%
   V\sim \text{Bernoulli}(\beta) 
$ %
 such that
\begin{align}
    L=T\oplus V
\end{align}  
\end{subequations}
where $T$ and $V$ are statistically independent,
for $\beta=\frac{q_{\mathsf{Z}}-\alpha}{1-2\alpha}$.
Notice that $\beta$ satisfies
$(1-\alpha)\beta+\alpha (1-\beta)=q_{\mathsf{Z}}$.

We begin with a construction of quantum encoding and decoding maps, along with a classical codebook. 
\paragraph{Quantum Code}
Recall that $\alpha$ is a parameter that satisfies \eqref{Equation:h2alpha}.
Let $(\mathcal{E}_n,\mathcal{D}_n)$ be a sequence of quantum encoder-decoder pairs that achieves the quantum capacity of the dephasing channel (without a relay),
\begin{align}
   \mathcal{Z}_\alpha(\rho)=(1-\alpha)\rho+\alpha\mathsf{Z}\rho\mathsf{Z} 
   \label{equation:D_alpha}
\end{align}
 for which a phase-flip occurs with probability  $\alpha$.

\paragraph{Classical Codebook}
Select $2^{nR_0}$ independent and identically distributed (i.i.d.) binary sequences
$v^n(s_0)$, for 
$s_0\in [2^{nR_0}]$, according to the
$\text{Bernoulli}(\beta)$ distribution.

The encoding and decoding procedures are described below.
\subsubsection{Encoding}
Apply the encoding map $\mathcal{E}_n$, and transmit. 

\subsubsection{Relay encoding}
Given the phase-flip sequence $\ell^n$:

Find a codeword 
$v^n(\widehat{s}_0)$ that is jointly typical with $\ell^n$.
    Based on the classical covering lemma \cite[Lemm. 3]{ElGamalKim:11b}, the probability that there exists an index 
    $\widehat{s}_0\in [2^{nR_0}]$ such that 
    $(v^n(\widehat{s}_0),\ell^n)\in T_\delta^{(n)}(p_{VL})$ tends to $1$ as $n\to\infty$, provided that 
    \begin{align}
        R_0&>I(V;L)\nonumber\\
        &=h_2(q_{\mathsf{Z}})-h_2(\alpha)
    \end{align}
which holds by \eqref{Equation:h2alpha}.
    If there is more than one such index, select the first.
    Then, transmit 
    $\widehat{s}_0$.

\subsubsection{Decoding}
Given the index
    $\widehat{s}_0$: 
\begin{enumerate}[(i)]
    \item   
    Apply phase-flips according to the pattern of the corresponding codeword, $v^n(\widehat{s}_0)$. That is, apply
\begin{align}
    \bigotimes_{i=1}^n \mathsf{Z}^{v_i(\widehat{s}_0)}
    \,.
\end{align}
Together with the noise from the channel, each qubit undergoes the error $\mathsf{Z}^{\ell_i}\cdot\mathsf{Z}^{ v_i(\widehat{s}_0)}=\mathsf{Z}^{\ell_i\oplus v_i(\widehat{s}_0)}$.
Hence, we deduce that the phase flips are governed by
\begin{align}
    t_i\equiv \ell_i\oplus v_i(\widehat{s}_0)\,.
\end{align}
Based on \eqref{Equations:T_V_L}, the probability distribution of $t_i$ is $\delta$-close to $\text{Bernoulli}(\alpha)$.

    \item Apply the decoding map $\mathcal{D}_n$.  
\end{enumerate}
After step 3-i, the effective channel is $\delta$-close to the dephasing channel $\mathcal{Z}_\alpha$, as in \eqref{equation:D_alpha}.
    Since $\delta$ is arbitrarily small, the coding scheme above guarantees reliable recovery of Alice's original state. 
    Furthermore, since the code sequence $(\mathcal{E}_n,\mathcal{D}_n)$ achieves the quantum capacity of the channel $\mathcal{Z}_\alpha$, the quantum coding rate approaches
    \begin{align}
        Q&=1-h_2(\alpha)\nonumber\\
        &=1-h_2(q_\mathsf{Z})+R_0-\varepsilon
    \end{align}
where the last equality follows from \eqref{Equation:h2alpha}.
    
\subsection{Entanglement-Assisted Communication}
\label{Subsection:EA_Achievability}
Consider the Pauli channel from Subsection~\ref{Subsection:Bell_Relay}, where Alice and Bob are provided with pre-shared entanglement assistance. 
As before, if $R_0\geq h_2(q_\mathsf{X})+h_2(q_\mathsf{Z})$, the result of 
$C_{\text{EA}}=2$
immediately follows from the noiseless superdense coding protocol.
Hence, assume %
$R_0= h_2(q_\mathsf{X})+h_2(q_\mathsf{Z})-2\delta$.

Thus,  there exists 
$\lambda\in [0,1]$ such that $\lambda R_0< h_2(q_\mathsf{X})$ and
$(1-\lambda) R_0< h_2(q_\mathsf{Z})$. Specifically, this holds for
$\lambda=\frac{h_2(q_\mathsf{X})-\delta}{R_0}$.
Similarly, there exist  $\alpha_{\mathsf{X}},\alpha_{\mathsf{Z}}\in \left[0,\frac{1}{2}\right)$ such that 
\begin{subequations}
    \label{Equation:h2alpha_Pauli}
\begin{align}
    h_2(\alpha_\mathsf{X})&= h_2(q_\mathsf{X})-\lambda R_0+\varepsilon 
    \label{Equation:h2alpha_X}
    \\
    h_2(\alpha_\mathsf{Z})&= h_2(q_\mathsf{Z})-(1-\lambda) R_0+\varepsilon 
    \label{Equation:h2alpha_Z}
\end{align}
\end{subequations}
and
\begin{subequations}
\begin{align}
    K&=T_\mathsf{X}\oplus V_\mathsf{X}\\
    L&=T_\mathsf{Z}\oplus V_\mathsf{Z}
\end{align}
\end{subequations}
for independent
$T_\mathsf{X}\sim \text{Bernoulli}(\alpha_\mathsf{X})$,
$V_\mathsf{X}\sim \text{Bernoulli}(\beta_\mathsf{X})$,
$T_\mathsf{Z}\sim \text{Bernoulli}(\alpha_\mathsf{Z})$, and
$V_\mathsf{Z}\sim \text{Bernoulli}(\beta_\mathsf{Z})$, and sufficiently small $\varepsilon>0$.

We use the same code construction as before, based on encoding and decoding maps $(\mathcal{E}_n,\mathcal{D}_n)$ for a Pauli channel where independent bit- and phase-flip errors occur with probability 
$\alpha_{\mathsf{X}}$ and $\alpha_{\mathsf{Z}}$, respectively. 

The relay can then find a compressed representation for $k^n$ and $\ell^n$, provided that 
\begin{subequations}
\begin{align}
        \lambda R_0&>I(V_\mathsf{X};K)\nonumber\\
        &=h_2(q_{\mathsf{X}})-h_2(\alpha_\mathsf{X})
        \intertext{and}
        (1-\lambda) R_0&>I(V_\mathsf{Z};L)\nonumber\\
        &=h_2(q_{\mathsf{Z}})-h_2(\alpha_\mathsf{Z}) \,.
    \end{align}
\end{subequations}
    Similarly, the decoder can then use those representations to simulate  a Pauli channel where  bit/phase-flips occur with probability 
$\alpha_{\mathsf{X}}$/$\alpha_{\mathsf{Z}}$. Therefore, the classical entanglement-assisted rate approaches
    \begin{align}
        R&=2-h_2(\alpha_\mathsf{X})-h_2(\alpha_\mathsf{Z})\nonumber\\
        &=2-h_2(q_\mathsf{X})-h_2(q_\mathsf{Z})+ R_0-2\varepsilon
    \end{align}
where the last equality follows from \eqref{Equation:h2alpha_Pauli}.

Now, consider the depolarizing channel. %
We assume $R_0< h_4\left(\frac{3q}{4}\right)$, since the result is immediate otherwise.
Then, there exists  $\alpha$ such that 
\begin{align}
    h_4\left(\frac{3\alpha}{4}\right)&= h_4\left(\frac{3q}{4}\right)- R_0+\varepsilon 
    \label{Equation:h4alpha}
\end{align}
and
\begin{subequations}
\begin{align}
    (L_\mathsf{X},L_\mathsf{Z})&=(T_\mathsf{X},T_\mathsf{Z})\oplus (V_\mathsf{X},V_\mathsf{Z})
\end{align}
\end{subequations}
 for independent pairs,
$(T_\mathsf{X},T_\mathsf{Z})\sim\left(1-\frac{3\alpha}{4}\,,\;\frac{\alpha}{4}\,,\;\frac{\alpha}{4}\,,\;\frac{\alpha}{4} \right)$ and 
$(V_\mathsf{X},V_\mathsf{Z})\sim\left(1-\frac{3\beta}{4}\,,\;\frac{\beta}{4}\,,\;\frac{\beta}{4}\,,\;\frac{\beta}{4} \right)$, $\beta=\frac{q-\alpha}{1-\alpha}$,
which yields
    \begin{align}
        R&=2-h_4\left(\frac{3\alpha}{4}\right)\nonumber\\
        &=2-h_4\left(\frac{3q}{4}\right)+ R_0-\varepsilon
    \end{align}
by \eqref{Equation:h4alpha}. This completes the achievability proof.

\section{Converse Proof}
\label{Section:Converse}
\subsection{Sending Quantum Information}
\label{Subsection:Dephasing_Converse}
Consider the quantum relay channel in Subsection~\ref{Subsection:Teleportation_Channel}.
Suppose that Alice prepares the entangled state
\begin{align}
    \ket{\Phi}_{MM_1}=\frac{1}{\sqrt{2^{nQ}}}
    \sum_{m=1}^{2^{nQ}} \ket{m}_M\otimes \ket{m}_{M_1}
\end{align}
and then uses the code in order to transmit $M_1$ to Bob.
Specifically, she applies an encoder
 $\mathcal{F}_{M_1\to A^n}$
that prepares the channel input state $\varphi_{MA^n}$, and transmits $A^n$ through the channel.
In addition, the relay transmits a classical message $S_0\in [2^{nR_0}]$ to the receiver.
Bob receives $S_0$ and $B^n$, and applies a decoder
$\mathcal{D}_{S_0 B^n\to \widehat{M}}$, which results in the final state $\rho_{M\widehat{M}}$. 

Now, consider a sequence of %
quantum codes that achieves an error $\gamma(n)$, where $\gamma(n)$
tends to zero as $n\to\infty$.
Hence, the final state $\rho_{M\widehat{M}}$ is $\gamma(n)$-close to the desired $\Phi_{MM_1}$.
We now show that 
\begin{align}
    Q
    &{\leq} \min\left\{1 \,,\; \frac{1}{n}I_c(\mathcal{Z}^{\otimes n})+R_0 \right\}+\varepsilon(n)
    \label{Eq:Q_Converse_NTS}
\end{align}
where $I_c(\mathcal{Z})$ is the coherent information of the dephasing channel $\mathcal{Z}(\rho)=(1-q_{\mathsf{Z}})\rho+q_{\mathsf{Z}} \mathsf{Z}\rho \mathsf{Z}$, without the relay (see \cite[Def. 13.5.1]{Wilde:17b}), for $\varepsilon(n)$ that tends to zero as $n\to\infty$. 

Consider that
\begin{align}
    nQ&=I(M\rangle M_1)_\Phi \nonumber\\
    &\stackrel{(a)}{\leq} I(M\rangle \widehat{M})_\rho+n\varepsilon(n) \nonumber\\
    &\stackrel{(b)}{\leq} I(M\rangle B^n S_0)_\rho+n\varepsilon(n)
\label{Eq:Converse_1}
\end{align}
where $(a)$ follows from entropy continuity \cite{Winter:16p}, for $\varepsilon(n)=2\gamma(n)Q+\frac{1}{n}(1+\gamma(n))$, %
and $(b)$ from the data processing inequality (DPI) for the coherent information \cite[Th. 11.9.3]{Wilde:17b}.

To show the first bound in \eqref{Eq:Q_Converse_NTS}, we apply the DPI again: 
\begin{align}
    nQ
    &{\leq} I(M\rangle A^n)_\varphi+n\varepsilon(n)\nonumber\\
    &{\leq} \log\mathrm{dim}(
    \mathcal{H}_A^{\otimes n})+n\varepsilon(n)\nonumber\\
    &{=} (1+\varepsilon(n))n
\label{Eq:Converse_2}
\end{align}
since $I(A\rangle B)_\rho \leq H(B)_\rho\leq \log \mathrm{dim}(\mathcal{H}_B)$ in general.
Hence, $Q\leq 1+\varepsilon(n)$.

As for the second bound, we observe that
\begin{align}
    H(M|B^n S_0)_\rho \geq H(M|B^n)_\rho-nR_0
\label{Eq:Converse_R0}
\end{align}
by the chain rule. To see this, notice that 
$H(M|B^n)_\rho-H(M|B^n S_0)_\rho=I(S_0;M|B^n)_\rho\leq \log(2^{nR_0})=nR_0$, as $S_0\in [2^{nR_0}]$ is a classical variable.
Thus, by \eqref{Eq:Converse_1} and \eqref{Eq:Converse_R0},
\begin{align}
    Q&\leq \frac{1}{n} I(M\rangle B^n)_\rho+R_0+\varepsilon(n)\\
    &\leq \frac{1}{n} I_c(\mathcal{Z}^{\otimes n})+R_0+\varepsilon(n)\,.
\end{align}
The converse part follows since 
$I_c(\mathcal{Z}^{\otimes n})=n[1-h_2(q_{\mathsf{Z}})]$ for the dephasing channel \cite{DevetakShor:05p}. %

\subsection{Entanglement-Assisted Communication}
\label{Subsection:EA_Converse}
We focus on the depolarizing channel from Subsection~\ref{Subsection:Bell_Relay}, where Alice and Bob are provided with pre-shared entanglement assistance.
Suppose that Alice stores a message in two classical registers, $M$ and $M_1$:
\begin{align}
    \pi_{MM_1}=\frac{1}{{2^{nR}}}
    \sum_{m=1}^{2^{nR}} \ketbra{m}_M\otimes \ketbra{m}_{M_1} \,.
\end{align}
In addition, Alice and Bob share an  an entanglement resource 
$\ket{\Psi}_{T_A T_B}$.
Alice applies an encoder
 $\mathcal{F}_{T_A M_1\to A^n}$ on $\pi\otimes \Psi$
to prepare the channel input state $\varphi_{M A^n T_B}$, and transmits $A^n$ through the channel.
In addition, the relay transmits a classical message $S_0\in [2^{nR_0}]$ to the receiver.
Bob receives $S_0$ and $B^n$, and estimates Alice's message through  a measurement
$\mathcal{D}_{S_0 B^n T_B\to \widehat{M}}$.

Now, consider a sequence of %
entanglement-assisted classical codes that achieves an error probability  that
tends to zero as $n\to\infty$.
By Fano's inequality and the DPI,
\begin{align}
    nR&\leq I(M;B^n T_B S_0)_\rho+n\varepsilon(n)
\end{align}
(see \cite[Th. 11.9.4]{Wilde:17b}).
We now show that 
\begin{align}
    R
    &{\leq} \min\left\{2 \,,\; I(\mathfrak{D}) +R_0 \right\}+\varepsilon(n)
    \,,
\end{align}
where $I(\mathfrak{D})$ is the mutual information of the depolarizing channel, without the relay (see \cite[Th. 21.3.1]{Wilde:17b}). 
The first bound follows by applying the DPI a second time: 
$I(M;B^n T_B S_0)_\rho\leq I(M;A^n T_B)_\varphi=I(M;A^n| T_B)_\varphi\leq 2\log \mathrm{dim}(\mathcal{H}_A^{\otimes n})=2n$, by \cite[Ex. 11.6.3]{Wilde:17b}. As for the second bound, by the chain rule, 
\begin{align}
    I(M;B^n  S_0T_B)_\rho
\leq I(M;B^n T_B )_\rho+nR_0
\end{align}
as $S_0\in [2^{nR_0}]$ is  classical  (see Property 11.7.1 and Theorem 11.5.1 in \cite{Wilde:17b}).
The converse part follows since the first term on the right-hand side is bounded by $nI(\mathfrak{D})$ based on standard arguments \cite[Eq. (21.103)-(21.108)]{Wilde:17b},
and 
$I(\mathfrak{D})=2-h_4\left(\frac{3q}{4}\right)$ for the depolarizing channel  (without a relay) \cite{BennettShorSmolin:99p}. %
This completes the proof. 
\qed

\section{Summary and Discussion}
\label{Section:Summary}
We introduced two quantum relay-channel models in which the relay has classical side information about Pauli errors affecting the transmitted quantum system and communicates with the destination through an orthogonal rate-limited link. The two models are motivated by quantum teleportation and superdense coding, %
and admit closed-form capacity formulas. %

For the dephasing relay channel, we showed that the quantum capacity is
\begin{align}
    C_{\text Q}=\min\left\{1 \,,\; 1-h_2(q_Z)+R_0 \right\} \,.
\end{align}
For the Pauli relay channel with independent bit- and phase-flip errors, the entanglement-assisted classical capacity is
\begin{align}
    C_{\text{EA}}=\min\left\{2\,,\;2-h_2(q_{\mathsf X})-h_2(q_{\mathsf Z})+R_0 \right\}\,,
\end{align}
with the analogous expression
\begin{align}
    C_{\text{EA}}=\min\left\{2\,,\;2-h_4\left(\frac{3q}{4}\right)+R_0 \right\}\,
\end{align}
for depolarizing noise.

A common feature of these results is that the relay's rate translates directly into communication rate: each additional bit per channel use available on the relay link increases the corresponding capacity by one bit, until the noiseless limit is reached. Operationally, the relay uses its limited communication resource to provide a compressed description of the Pauli-error sequence. The destination uses this description to partially cancel the errors, effectively converting the original channel into one with reduced noise. Thus, the optimal strategy has a natural compress-forward interpretation, with the relay compressing information about the channel noise rather than decoding the transmitted message.
As can be seen in Figures~\ref{Figure:Dephasing_Capacity_R0}-\ref{Figure:Depolarizing_Capacity_R0},
the capacity increases linearly with the relay's rate until the corresponding noiseless teleportation or superdense-coding limit is attained. 

These models illustrate a setting in which the benefit of classical side information in a quantum communication network can be characterized exactly. They also suggest a broader question: to what extent can rate-limited information about the channel environment increase quantum communication rates for more general noise models? Extending the present approach beyond Pauli channels, as well as allowing the relay to receive and process quantum side information, may provide a path toward capacity results for more general quantum relay channels.

\begin{figure}
    \centering

  \begin{tikzpicture}
    \begin{axis}[
        width=5.75cm,       %
        height=4.5cm,      %
      xlabel={$R_0$},
      ylabel={$C_{\text{Q}}$},
      ylabel style={rotate=-90}, %
      xmin=0, xmax=1.65,
      ymin=0, ymax=1.1,
      xtick distance=0.5,
      ytick distance=0.25,
      grid=both,
      grid style={dashed, gray!30},
      axis lines=left,
      declare function={
        q = 0.15; %
        h2(\x) = (\x <= 0 ? 0 : -\x*(ln(\x)/ln(2)) - (1-\x)*(ln(1-\x)/ln(2)));
        f(\x) = min(1, 1 - h2(q) + \x);
      }
    ]
      \addplot[
        domain=0.00001:1.6,
        samples=600,
        thick,
        red
      ] {f(x)};
    \end{axis}
  \end{tikzpicture}

    \caption{Quantum capacity for the dephasing relay channel as a function of the relay's rate limit $R_0$ (in bits per channel use), for a dephasing probability of $q_{\mathsf{Z}}=0.15$. }
    \label{Figure:Dephasing_Capacity_R0}
\end{figure}
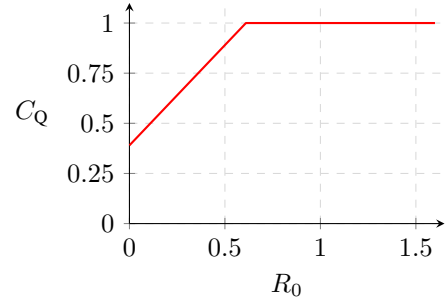
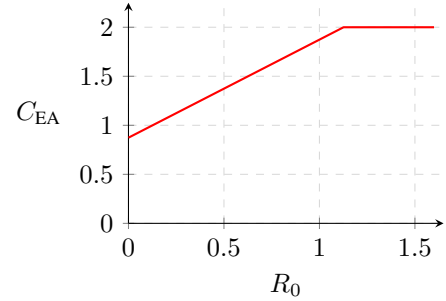
\begin{figure}
    \centering

  \begin{tikzpicture}
    \begin{axis}[
        width=5.75cm,       %
        height=4.5cm,      %
      xlabel={$R_0$},
      ylabel={$C_{\text{EA}}$},
      ylabel style={rotate=-90}, %
      xmin=0, xmax=1.65,
      ymin=0, ymax=2.25,
      xtick distance=0.5,
      ytick distance=0.5,
      grid=both,
      grid style={dashed, gray!30},
      axis lines=left,
      declare function={
        q = 0.3; %
        h4q(\x) = (\x <= 0 ? 0 : -(1-0.75*\x)*(ln(1-0.75*\x)/ln(2)) - 0.75*\x*(ln(\x/4)/ln(2)));
        f(\x) = min(2, 2 - h4q(q) + \x);
      }
    ]
      \addplot[
        domain=0:1.6,
        samples=600,
        thick,
        red
      ] {f(x)};
    \end{axis}
  \end{tikzpicture}

    \caption{Entanglement-assisted classical capacity for the depolarizing relay channel as a function of the relay's rate limit $R_0$, for a probability of $q=0.3$ for depolarization. }
    \label{Figure:Depolarizing_Capacity_R0}
\end{figure}

\section*{Acknowledgments}
We %
thank Gerhard Kramer (TUM)
for useful discussions.
Yilin and Pereg were supported by  ISF n. 939/23 and 2691/23,
DFG %
 n. 2032991, Technion OMC 
 n. 86160946,   
 IIA %
 Post-Quantum Communications n. 2035868,
  VATAT for QERNEL Quantum  Computing Research Hub n.
2072651, and the  HD Quantum Center. %

\begin{appendices} %
{

}
\end{appendices}

{

\balance

\bibliography{ITW_2027/references_ITW}%

}

\end{document}